\documentclass{article}
\usepackage{graphicx,amssymb,amsmath,listings,url}

\title{Divide-and-Conquer 3D Convex Hulls on the GPU}

\author{
Jeffrey M. White \thanks{Department of Computer Science,
California State University, Fullerton, {\tt jeffreymarkwhite@gmail.com}}
\and
Kevin A. Wortman \thanks{Department of Computer Science,
California State University, Fullerton, {\tt kwortman@fullerton.edu}}
}
        
\index{White, Jeffrey}
\index{Wortman, Kevin}

\begin{document}
\thispagestyle{empty}
\maketitle

\begin{abstract}
We describe a pure divide-and-conquer parallel algorithm for computing 3D convex hulls. We implement that algorithm on GPU hardware, and find a significant speedup over comparable CPU implementations.
\end{abstract}

\section{Introduction}

The \emph{3D convex hull problem} is to identify, for a given set of $n$ points in $\mathbb{R}^3$, the minimal set of input points such that the convex envelope of those points contains all input points. The problem is fundamental to computational geometry and has been studied extensively. Several $O(n \log n)$ time algorithms are known, with various trade-offs in constant factors, simplicity, numerical robustness, data structure dependencies, and nondegeneracy requirements  (see e.g. \cite{Akl1978219} \cite{Barber:1996:QAC:235815.235821} \cite{springerlink:10.1007/BF02573985} \cite{Eddy:1977:NCH:355759.355766}  \cite{kirkpatrick:287} \cite{Preparata:1977:CHF:359423.359430} \cite{seidel81}). Chan's celebrated output-sensitive algorithm \cite{springerlink:10.1007/BF02712873} runs in $O(n \log h)$ time, where $h$ denotes the number of faces in the output hull, which is asymptotically optimal.

A \emph{graphics processing unit (GPU)} is a parallel coprocessor available in commodity computers. An outgrowth of the computer gaming industry, GPUs utilize a highly-parallel single instruction multiple data (SIMD) architecture. At a high-level, GPUs work by applying a concise constant-space function called a \emph{kernel} to all elements of an array simultaneously. Kernels are written in \emph{domain specific embedded languages (DSELs)} such as NVIDIA's CUDA \cite{CUDAGuide2.0} or the OpenCL \cite{Opencl_theopencl} open standard. Each kernel instance is passed an integer \emph{global identifier (id)} which is customarily used to delineate the ranges of input that each kernel invocation applies to. The potential performance, measured in either gigaFLOPS or memory bandwidth, of GPUs is substantially greater than that of multicore CPUs. However, realizing this potential on practical problems, besides the embarrasingly-parallel graphics applications for which GPUs were originally designed, has proven challenging. By and large, existing parallel algorithms depend on facilities, such as message passing and/or synchronization primitives, which are unavailable in the GPU environment. Yet, GPUs are purpose-built for high performance computation on low-dimensional geometric objects, and the opportunity to apply them to computational geometry problems cannot be ignored.

While the 3D convex hull problem has been studied extensively in the standard computational model, precious little past work is applicable to GPU implementations. As stated above, GPU kernels cannot communicate with or synchronize against each other. This limitation rendered unusable every PRAM-model algorithm we surveyed (e.g. \cite{pram_alg}). Further, running kernels have no provision for dynamic memory; their collective input and output must be allocated before the kernels execute \emph{en masse} and freed afterward. Accordingly dynamic data structures are off limits. The absence of the doubly connected edge list (DCEL) structure is a particularly formidable obstacle in this context.

There are several results on computing 2D hulls on the GPU  \cite{JurkiewiczDanilewski2011} \cite{ro08cuda} \cite{5763404}, but results on the more general and complex 3D problem have been elusive. While preparing this manuscript, we became aware of an independent result on the 3D problem \cite{ghull}. That algorithm uses heuristics to cull many, but not all, interior points on the GPU, then feeds the remaining points to a black-box CPU hull implementation (e.g. QuickHull \cite{Barber:1996:QAC:235815.235821}). The algorithm presented here achieves competitive performance using a pure GPU divide-and-conquer approach, whose worst case running time is not impacted by the presence of outlier points, and which is conceptually simpler.

\section{Algorithm}

\begin{figure}
\centering
\includegraphics[trim = 2mm 2mm 2mm 2mm, clip, width=90mm]{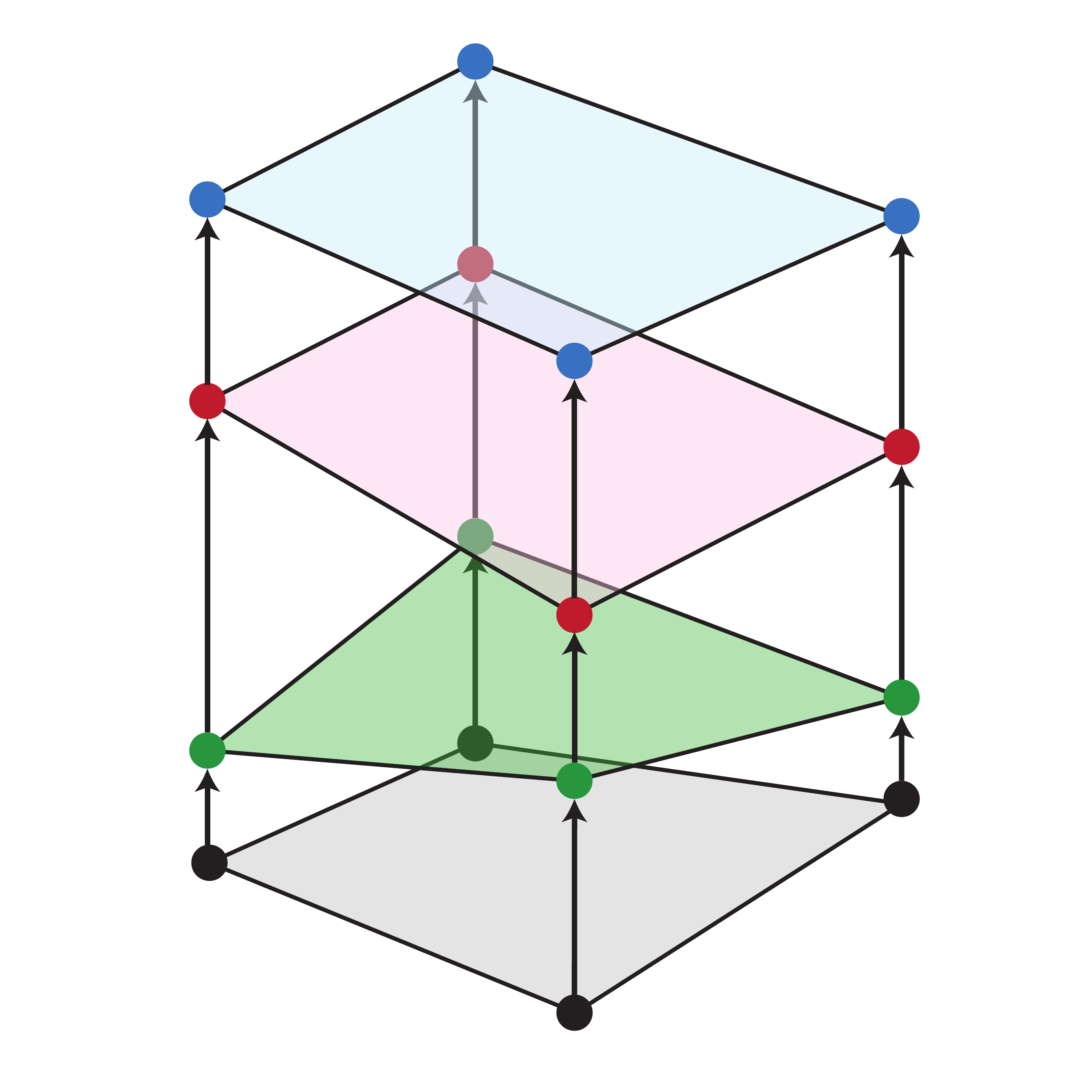}
\caption{Algorithm Events.}
\label{algorithm_events}
\end{figure}

Our algorithm is an adaptation of Chan's \emph{minimalist} 3D convex hull algorithm \cite{chan_minimalist}. Note that this $O(n \log n)$-time algorithm is distinct from the $O(n \log h)$-time algorithm mentioned earlier, also authored by Chan. The minimalist algorithm is, by design, a straightforward top-down divide-and-conquer algorithm for computing 3D convex hulls. It was originally motivated by pedagogical needs for an algorithm that achieves a favorable $O(n \log n)$ running time, while being simple to explain and implement and avoiding dependency on difficult data structures or algorithms. Serendipitously these design constraints correspond to those imposed by the GPU.

The minimalist algorithm works by recasting the 3D problem as a 2D \emph{kinetic} problem. 3D $(x, y, z)$ points are mapped to $(x, y, \Delta y)$ points with an initial $(x, y)$ starting point and $\Delta y$ vertical rate of speed. As time $t$ advances, the points move at distinct velocities, which triggers structural changes in the convex hull of the points (see Figure 1). Computing the convex hull of the original 3D points may be visualized as computing a \emph{kinetic movie} of these configurations for all values $-\infty < t < \infty$. The algorithm represents this movie as a chronological sequence of \emph{events} when input points are added to, or removed from, the hull. Input points are presorted by $x$-coordinate; then event sequences for roughly equal-size subsets are recursively generated, then combined by a Graham-scan-like $O(n)$ merging process. In the base case a single point nominates itself as the only convex hull point.

While the minimalist algorithm boasts many of the features necessary for GPU implementation, it cannot be ported to the GPU directly. GPU kernels cannot be recursive, so the top-down divide-and-conquer approach is inappropriate. Instead, the algorithm must be reoriented into one or more mapping steps where an array of input data elements are mapped by a kernel to an array of output data elements. We achieve this reorientation by rewriting the minimalist algorithm to use \emph{bottom-up} divide and conquer. We define a \emph{movie array} data structure as a table of event logs. Our algorithm allocates a single movie array, and initializes one trivial event log for each input point. Then, our algorithm repeats a \emph{merge step} that combines each pair of event logs with adjacent indices into a single event log. A merge step maps a movie array with $n$ logs of length at most $l$ to a new array with at most $\lceil n/2 \rceil$ logs of length at most $2l$ each. Thus, after $\lceil \log_2 n \rceil$ merge steps, the movie array contains a single event log for the entire point set. The key property of this algorithm with respect to GPU computation is that each log merge may be performed entirely independently of the others. Each kernel has a particular range of input movie array indices to read from, and a corresponding range of output indices to write to, and may perform its computation independently of other concurrent kernel instances.

\section{Implementation}

Our implementation of the GPU algorithm follows the bottom-up divide-and-conquer design as mentioned above. As shown in Figure \ref{point_lists}, the point structure in the CPU algorithm uses a doubly linked list connected by pointers. The idea is to divide the sorted list down into trivial subsequences and build the list back up to the desired set of faces on the convex hull. Memory pointers are difficult (though not impossible) to move between the CPU and GPU since the two devices have distinct memory spaces. Also, on the GPU each kernel instance needs to seek to its assigned sub-input based on its global id, which could take $O(n)$ time using a list structure. For these reasons, our GPU implementation uses arrayed lists with integer indices rather than linked lists with node addresses (Figure \ref{point_lists}).

\begin{figure}
\begin{lstlisting}
// CPU Algorithm Point
struct Point {
  double x, y, z;
  Point *prev, *next;
  void act() {...}
};

// GPU Algorithm Point
struct Point { 
  cl_float x;
  cl_float y;
  cl_float z;  
  cl_int prev;
  cl_int next;
};
\end{lstlisting}
\caption{Differences in the Point datatype.}
\label{point_datatypes}
\end{figure}

\begin{figure}
\begin{lstlisting}
// CPU Algorithm list of points
Point *P = new Point[n];
...
// Sorts points into a doubly 
// linked list based x-coordinate.
Point *list = sort(P, n);

// event lists
Point **A = new Point *[2*n];
Point **B = new Point *[2*n];


// GPU Algorithm list of points
Point *P = (Point *) 
    malloc(n*sizeof(Point));

// event lists
cl_int *A = (cl_int *) 
    malloc(2*n*sizeof(cl_int));
cl_int *B = (cl_int *) 
    malloc(2*n*sizeof(cl_int));
\end{lstlisting}
\caption{Differences in list creation.}
\label{point_lists}
\end{figure}

Modifying the way data is stored impacts the way data is accessed. Figure \ref{act_functions} shows the differences in \texttt{act()} function used for inserting and deleting points from event logs. Figure \ref{time_calculations} shows the differences in passing potential faces into the event-time calculations.

\begin{figure}
\begin{lstlisting}
// CPU Algorithm act() function call
point->act()

// CPU Algorithm act() function
struct Point {
...
  void act() {  
    if (prev->next != this) {
      // insert point 
      prev->next = next->prev = this;
    }
    else { 
      // delete point
      prev->next = next; 
      next->prev = prev; 
    }
  }
};

// GPU Algorithm act() function call
act(pointIndex);

// GPU Algorithm act() function
void act(int pointIndex) {
  if (P[P[pointIndex].prev].next 
      != pointIndex) {   
    // insert point
    P[P[pointIndex].prev].next 
    = P[P[pointIndex].next].prev 
    = pointIndex;
  }
  else { 
    // delete point
    P[P[pointIndex].prev].next 
    = P[pointIndex].next;
    P[P[pointIndex].next].prev 
    = P[pointIndex].prev;
  }	
}	
\end{lstlisting}	
\caption{Differences in \texttt{act()} functions.}
\label{act_functions}
\end{figure}

\begin{figure}
\begin{lstlisting}
// CPU Algorithm time[0] calculation
t[0] = time(B[i]->prev, 
            B[i], 
            B[i]->next);

// GPU Algorithm time[0] calculation
t[0] = time(P[B[i]].prev, 
            B[i], 
            P[B[i]].next);
\end{lstlisting}
\caption{Differences in time calculations.}
\label{time_calculations}
\end{figure}

The implementation process began with converting the original CPU algorithm to use arrays rather then pointers to represent the data. Point data is implemented as its own data type with the $x, y,$ and $z$ values along with indices to represent the next and previous pointers to reference other points based on their array index. Also, instead of having two pointer lists, $A$ and $B$, we have two arrays of indices that reference a master list $P$ of points. 

\begin{figure}
\begin{lstlisting}
dataOffsetValue = 2;
totalMergesLeft = numberOfPoints/2;
do {
    numberOfThreads = totalMergesLeft;
    runGPUkernels();
    swap(A, B);
    dataOffsetValue = dataOffsetValue*2;
    totalMergesLeft = totalMergesLeft/2;
} while(totalMergesLeft > 1);
\end{lstlisting}
\caption{Main outer loop ran on the CPU to handle the execution of threads on the GPU.}
\label{main_loop_figure}
\end{figure}

Another significant change we made to the design is the conversion from a top-down design to a bottom-up design. Instead of using recursion, the heart of the algorithm is placed within one \textbf{while} loop as shown in Figure \ref{main_loop_figure}. Before implementing this routine as OpenCL kernel code, we wrote a simulation to run on the serial CPU to ensure validity of the algorithm. The ultimate goal of writing a simulation is to avoid the troublesome task of debugging GPU kernel code. This simplified the task of converting the simulation code to GPU kernel code and required only minimal modifications.

Figure \ref{main_loop_figure} shows pseudocode for the main outer loop which runs on the CPU. The main loop uses two movie array structures, both of which exist on the GPU. The two structures alternate between serving as the input and output of a merge step. This approach makes it possible to avoid transferring point data between the GPU and CPU inside the loop, which is desirable as that is an expensive operation. The \texttt{dataOffsetValue} is used to calculate the location of where the head of the \texttt{leftGroupIndex} and \texttt{rightGroupIndex} exist on the globally accessed master list of points $P$ as shown in Figure \ref{kernel_variables}. To handle the way the CPU algorithm swaps lists $A$ and $B$ in each divide routine, we swap the kernel arguments of $A$ and $B$ in the \texttt{swap(A, B)} function after each iteration of merges. Following the \texttt{swap(A, B)} function, \texttt{dataOffsetValue} is updated to tie into the next set of group index calculations. Finally, \texttt{totalMergesLeft} is cut in half to represent the number threads to take place in the next iteration of merges. When \texttt{totalMergesLeft} reaches less than 2, the algorithm exits the main \textbf{while} loop as there is no pair of hulls left to be merged together; only one hull is left which represents the final solution.

\begin{figure}
\begin{lstlisting}
// the index of where the head of the 
// left group of the list can be found 
// on the globally accessed array
leftGroupIndex 
 = global_ID*dataOffsetValue;

// the index of where the head of the 
// right group of the list can be found 
// on the globally accessed array
rightGroupIndex 
 = [leftGroupIndex+((global_ID+1) 
   *dataOffsetValue)]/2;

// the index of where the globally
// accessed event list begins for the
// group of merges based on the global_ID
eventListOffset = leftGroupIndex*2;
\end{lstlisting}
\caption{GPU kernel code: how the GPU knows which hulls should be merged and which parts of the global data to access.}
\label{kernel_variables}
\end{figure}


\section{Experimental Results}

The GPU algorithm shows significant improvements over the CPU algorithm. Peak performance of the GPU algorithm reaches close to a 6x speedup over the CPU algorithm. Figures \ref{chart_figure} and \ref{graph_figure} illustrate the runtime of both algorithms in milliseconds.

The CPU algorithm runtime calculations are based on a Intel \textsuperscript{\textregistered} Core\textsuperscript{TM} i3-2330M Processor with 2 cores capable of processing 2 threads each at a rate 2.2 gigahertz, and acheives about 25.0 gigaFLOPS according to the LINPACK benchmark tool. The GPU algorithm runtime calculations are based on an ATI Radeon HD 6470m graphics card with 32 stream cores each with 5 processing elements capable of pipelining data at a rate of 700-750 megahertz. Peak performance for this GPU can potentially reach 224-240 gigaFLOPS \cite{amdspecs}. Both the CPU and GPU, as described, rank low on the spectrum of hardware available based on performance.

\begin{figure}
\centering
  \begin{tabular}{|l|l|l|}
    \hline
    Points & GPU Alg. (ms) & CPU Alg. (ms) \\ \hline
    4 & 0 & 0 \\ 
    8 & 0 & 0 \\ 
    16 & 0 & 0 \\ 
    32 & 0 & 0 \\ 
    64 & 0 & 0 \\ 
    128 & 0 & 1 \\ 
    256 & 0 & 3 \\ 
    512 & 0 & 7 \\ 
    1024 & 0 & 9 \\ 
    2048 & 1 & 10 \\ 
    4096 & 2 & 13 \\ 
    8192 & 4 & 25 \\ 
    16384 & 9 & 34 \\ 
    32768 & 23 & 59 \\ 
    65536 & 38 & 123 \\
    131072 & 56 & 233 \\ 
    262144 & 88 & 465 \\ 
    524288 & 173 & 941 \\
    1048576 & 354 & 1869 \\ 
    2097152 & 652 & 3732 \\ 
    4194304 & 1309 & 7494 \\ 
    8388608 & 2598 & 15047 \\
    \hline
  \end{tabular}
\caption{Run time for $n$ data points.}
\label{chart_figure}
\end{figure}

\begin{figure}
\centering
\includegraphics[trim = 20mm 83mm 20mm 83mm, clip, width=90mm]{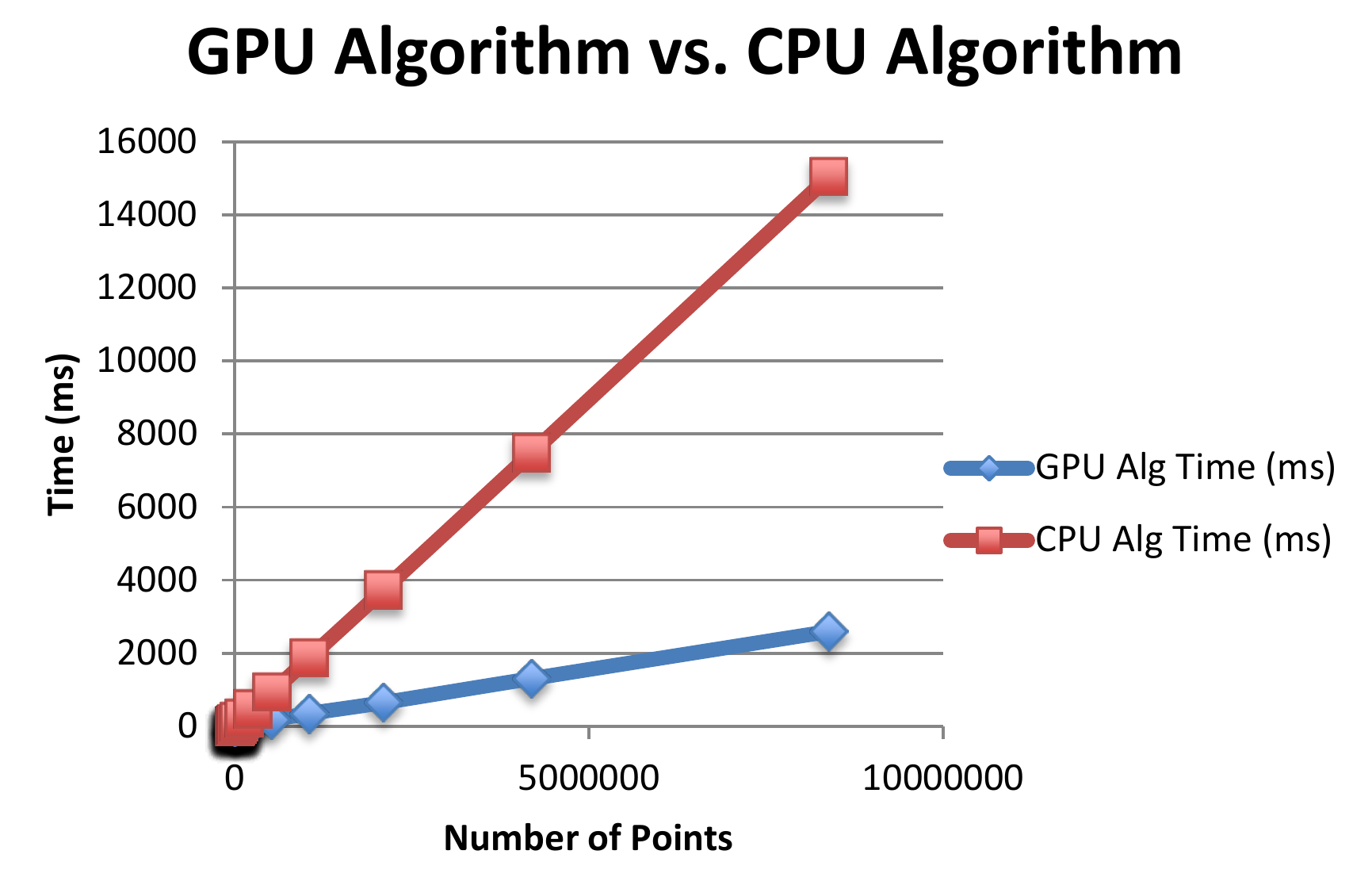}
\caption{Run time graph for $n$ data points.}
\label{graph_figure}
\end{figure}

Originally, a hybrid approach to the GPU algorithm seemed to be a more attractive solution to solving the problem. The hybrid GPU algorithm would perform nearly all of the merge steps on the GPU, then perform the last few steps on the CPU after the \texttt{totalMergesLeft} variable reached a certain value. The premise of this approach is that the last few iterations are poorly parallelizable and could be more quickly performed by a serial CPU. To accomplish this, the partially computed data would need to be copied from GPU memory to memory that the CPU has access to. On the CPU side, there would be a similar algorithm which would finish the rest of the computation using that same bottom-up style algorithm.

Surprisingly, our experimental results showed that those last few merge iterations take an insignificant amount of time -- less than one millisecond. So the hybrid approach is overly-complex, and implementing it would have been an instance of \emph{premature optimization.} The final design of the GPU algorithm takes place entirely on the GPU rather then on both GPU and CPU hardware. The GPU algorithm just requires the use of the CPU for the required OpenCL setup routines and ultimately to read in the data and output the data; the GPU completes all the extensive computations. 

Something we found interesting is the ratio of speedup improvements over the CPU algorithm as the data set increases. For a data set of four points, the speed up is close to 6x. As the data set approaches 32,768 points, the speedup decreases to about 2.5x. From 32,768 points and on, the speedup increases back to about 6x.

Our roughly 6x speedup is notable since it approaches the maximum potential improvement achievable on our hardware. According to the manufacturers, our GPU is capable of roughly 9 times more gigaFLOPS than our CPU. So the greatest conceivable speedup factor is roughly 9, which would correspond to an embarrasingly-parallel problem with negligible overhead. Our implementation comes close to realizing this full potential despite the obstacles inherent in parallelizing the 3D convex hull problem.

\section{Conclusion}

We have shown that bottom-up adaptation of the minimalist divide-and-conquer algorithm for 3D convex hulls is fast, practical, and reasonably straightforward. The approach achieves run-time performance comparable to past GPU implementations of convex hull algorithms.

In performing this exercise, we did make two counterintuitive conclusions. First, while OpenCL and CUDA are intended to be high-level abstractions of GPU hardware, we nonetheless faced many obstacles related to low-level concerns such as memory management, memory hierarchies, and thread scheduling. Second, our intuition was that the overhead of starting and scheduling kernel applications would become a major bottleneck in the later steps of the algorithm.  However, empirical results demonstrated this to be a non-issue.

The following are potential areas for future work:
\begin{itemize}
\item Higher-level libraries or tools for implementing divide-and-conquer algorithms on the GPU.
\item A suite of compatible, parallel GPU implementations of fundamental computational geometry algorithms.
\item In particular, an arrangement data structure, e.g. doubly connected edge list, is a prerequisite to implementing many well-motivated algorithms.
\end{itemize}


\small 
\bibliographystyle{abbrv}

\bibliography{hull3dpaper}


\end{document}